\documentstyle[twoside,fleqn,epsf,espcrc2]{article}

\newcommand{\aSF}    [0]{\alpha_{\scriptscriptstyle SF}}
\newcommand{\aMS}    [0]{\alpha_{\scriptscriptstyle \overline{MS}}}
\newcommand{\ao}     [0]{\alpha_{0}}
\newcommand{\aot}    [0]{\tilde{\alpha}_{0}}
\newcommand{\gqSF}   [0]{g_{\scriptscriptstyle SF}^2}
\newcommand{\btwoeff}[0]{b_2^{\scriptscriptstyle eff}}
\newcommand{\btwoSF} [0]{b_2^{\scriptscriptstyle SF}}
\newcommand{\btwoMS} [0]{b_2^{\scriptscriptstyle \overline{MS}}}
\newcommand{\myref}  [4]{\bibitem{#1} #3, #4}
\title{Two loop expansion of the Schr\"odinger functional
coupling $\aSF$ in $SU(3)$ lattice gauge theory}

\author{A.~Bode\address{Institut f\"ur Physik,
			Humboldt-Universit\"at zu Berlin, \\
                        Invalidenstr. 110,
                        D-10115 Berlin, Germany}%
               \thanks{Supported by Deutsche Forschung Gemeinschaft, 
		        Grant No. Wo 389/3-2}}
       
\begin{document}

\begin{abstract}
The two loop coefficient of the expansion of the Schr\"odinger
functional coupling in terms of the lattice coupling is calculated for
the $SU(3)$ Yang-Mills theory. This coefficient is required to relate
lattice data to the $\overline{MS}$-coupling.  As a byproduct of the
calculation, the Schr\"odinger functional is improved to two loop
order and the three loop coefficient of the beta function in the
SF-scheme is derived.
\end{abstract}

\maketitle

\section{Introduction}
In the framework of the Schr\"odinger functional the renormalized
coupling $\aSF$ can be traced from low to high energy numerically on
the lattice with finite size techniques. At high energy a conversion
to $\aMS$ with perturbation theory is possible. The relation can be
calculated by expanding both $\aMS$ and $\aSF$ in $g_0$, the lattice
bare coupling.

This program is completed for $SU(2)$
\cite{SU2_ASF-G0,SUN_AMS-G0_LETTER,SU2_NUMERIC_2} where the two loop
relation between $\aMS$ and $\aSF$ was required in order to avoid a
significant source of error in the conversion to $\aMS$. In the
$SU(3)$ case \cite{SU3_NUMERIC} the remaining step is to compute the two
loop coefficient between $\aSF$ and $g_0$.

This coefficient also enters into an extension of the program to
include quenched quarks, which is discussed in Martin L\"uschers contribution
to these proceedings.

The calculation is analogous to the $SU(2)$ case %group
\cite{SU2_ASF-G0}. We have to calculate the perturbative coefficients
depending on the box size $I=L/a$, where
$\aSF$ is defined. From this we are able to extract the continuum
results and the $O(a)$ improvement terms. Combining the numerical data
and the two loop relations results in the expected error reduction
comparable to the $SU(2)$ case.
\section{Definition of $\aSF$}
The coupling $\aSF(q)=\gqSF(L)/(4\pi)$, at $q=L^{-1}$ is defined via the
effective action $\Gamma$:
\begin{equation}
\exp(-\Gamma)=\int D[U] \exp(-S[U])
\end{equation}
where the Wilson action
\begin{equation}
S[U]=\frac{1}{g_0^2} \sum_{p} w(p) \mbox{tr}\{1-U(p)\}	
\end{equation}
is modified by the weight $w$ to achieve the $O(a)$
improvement: $w$ differs from one only for the plaquettes attached to
the boundary fields, where $w=c_t(g_0)$ is expanded in a series of
$g_0$. The coupling 
\begin{equation}
\gqSF(L)=\left.\frac{\Gamma_0'(L,\eta)}{\Gamma'(L,\eta)}\right|_{\eta=0} 
\end{equation}
is normalised via the classical action minimum $\Gamma_0$.
The derivative is with respect to $\eta$, which parametrises the
boundary fields applied at $t=0,L$. The remaining three space
dimensions have periodic boundary conditions. Details about the chosen
diagonal and constant boundary fields can be found in
\cite{SU3_NUMERIC}.

\section{Perturbative expansion and calculation}
The perturbative expansion has to be taken around the induced nonzero
background field. Since the propagators are not known analytically, we
calculate them numerically for each $I$ and sum them up within each of
the 13 diagrams\cite{SU3_ASF-G0[LONG]}. The use of symmetries
(for instance translation invariance and the cubic group
in space) in this context reduces the numerical effort. The
propagators are diagonal in the root basis of $SU(3)$ and in spatial
momentum space, and have been calculated there by a recurrence
relation in time. Numerical efficiency and preserving maximal
precision was taken into account in the program. In particular the
$\eta$ derivatives were taken analytically.

The expected independence of the gauge parameter $\lambda_0$ of
$\Gamma_2'$, 
the validity of the symmetries,
a comparison of $\Gamma_2'$ with a numerical derivative of $\Gamma_2$
and the recalculation of the $SU(2)$ results
are applied as tests as well as an independent calculation of Peter
Weisz for small $I$.

In 2 months CPU-time on a HP735 with 128MB main memory we obtained
$m_2^X(I)$ in the range $I=4 \ldots 32$, where:
\begin{eqnarray}
&&\gqSF(L)= g_0^2+m_1(I) g_0^4+m_2(I)g_0^6 \ldots\\
&&m_1(I)  = m_1^a(I)+c_t^{(1)} m_1^b(I) \\
&&m_2(I)  = m_2^a(I)+c_t^{(1)} m_2^b(I)  
               +\left[c_t^{(1)}\right]^2 m_2^c(I)  \\\nonumber && \qquad\qquad
             {}  +c_t^{(2)} m_2^d(I)+ m_1(I)^2
\end{eqnarray}
From \cite{PRIV-COMM-ML} $m_1^a$ is known and $m_1^b,m_2^c,m_2^d$ are
derived analytically.

Symanzik's analysis suggests as asymptotic expansion of $m_2^X$  
\begin{equation}
m^X_2(I)=\sum_{n=0}^{\infty} \frac{r^X_n+s^X_n \ln(I)+ t^X_n \ln^2(I)}{I^{n}}
\end{equation}
where we expect for instance $t^a_1=0$ from tree level improvement. 

Using the L\"uscher-Weisz \cite{BLOCKING} blocking method, which is based on
\begin{equation}
R_n[f](L)=f(L)+\frac{L}{n} \frac{\partial}{\partial L}f(L) \quad n\ge1
\end{equation}
With $f(L)=a+b\ln^m(L)/L^n$ we get:
\begin{equation}
R_n[f](L)=\left\{a+c/L^{n+1} \qquad \quad  m=0 
                      \atop a+c\ln^{m-1}(L)/L^n \quad m>0 \right. \;.
\end{equation}
This enables us to cancel some powers in the residual terms. We used a
symmetric lattice derivative and traced the numerical errors during
the blocking. The error of the constant term was obtained by a fit to
the residual terms. 

The universal $\beta$-function coefficient $b_1$ was subtracted after
confirming it within 1.3\%. 
At present all errors are very conservative. A more detailed error
analysis will be presented in \cite{SU3_ASF-G0[LONG]}.
The one loop improvement coefficient $c_t^{(1)}$ 
was confirmed within errors. This should be considered as
an additional test of the calculation.

Up to $O(I^{-2})$ times logarithms we get:
\begin{eqnarray} \label{m1}
&& m_1(I) = 2b_0\ln(I)+.368282(11) \\ \label{m2}
&& m_2(I) = 2b_1\ln(I)+.048085(63)+m_1(I)^2
\end{eqnarray}
The continuum results are obtained by formally neglecting all negative
powers of $I$, so eqn. (\ref{m1}),(\ref{m2}) are the continuum results.
\section{Applying the expansion}
Together with the expansion of $\aMS$ in $g_0$
\cite{SUN_AMS-G0_LETTER} we get:
\begin{eqnarray} \label{aMSaSF}
&&\aMS(sq)= \aSF \! +\!c_1(s) \aSF(q)^2\! +\!c_2(s) \aSF(q)^3 \\
&& c_1(s)= -8 \pi  b_0 \ln(s) + 1.25562(14)\\
&& c_2(s)= c_1(s)^2 - 32 \pi^2 b_1 \ln(s) +1.197(10)
\end{eqnarray}

With $\btwoMS$ \cite{SUN_B2MS} we are able to quote
$\btwoSF=0.4827(88)/(4 \pi)^3$, where a fit to the data of numerical
simulations gives \mbox{$\btwoeff=1.5(8)/(4 \pi)^3$}. There is no
reason that the coefficient of the fit coincides with $\btwoSF$, since
$\aSF$ is traced nonperturbativly on the
lattice. Nevertheless the order is the same and we expect as quoted in
\cite{SU3_NUMERIC} the same error reduction as in the $SU(2)$.

The perturbative relation (\ref{aMSaSF}) between $\aSF$ and $\aMS$
involves the scale factor $s$. Fixing them by demanding $c_1(s)=0$ or
$c_2(s)$ to be minimal gives comparable results:
\begin{eqnarray}
&& \aMS(   2.048q)    =\aSF(q)+   0.271(11)  \aSF(q)^3 \\
&& \aMS(   2.529q)    =\aSF(q)-   0.36895    \aSF(q)^2 \\ \nonumber
&& \phantom{\aMS(2.529q)=} {}+0.135(11)  \aSF(q)^3
\end{eqnarray}
where the quantities quoted without error are in all digits
significant. Note beside the scale factor the small coefficients.
This perturbative relation between the two physical couplings should
be safe for small enough couplings. Using the smallest numerically
determined coupling $\aSF$ we get for the $SU(3)$ gauge theory:
\begin{eqnarray}
&& \aMS(14.5\;\mbox{GeV}) = 0.1146(22)(2) \\
&& \aMS(78.2\;\mbox{GeV}) = 0.08407(121)(5) 
\end{eqnarray}
where the first error arises from the numerical uncertainties and the
scale determination at low energies. The second error results from the
residual evolution and the conversion to the $\overline{MS}$ scheme. 
This should be compared with the results where $c_2$ is unknown and
$\btwoeff$ is used:
\begin{eqnarray}
&& \aMS(14.5\;\mbox{GeV}) = 0.1145(23)(15) \phantom{00} \\
&& \aMS(78.2\;\mbox{GeV}) = 0.08380(121)(59)
\end{eqnarray}
The error reduction of the conversion by a factor of 10 was the
main goal of this calculation.\\
\begin{minipage}[t]{7.5cm}{ 
 \parbox[t]{0cm} 
{\epsfxsize=7.5cm \epsfbox{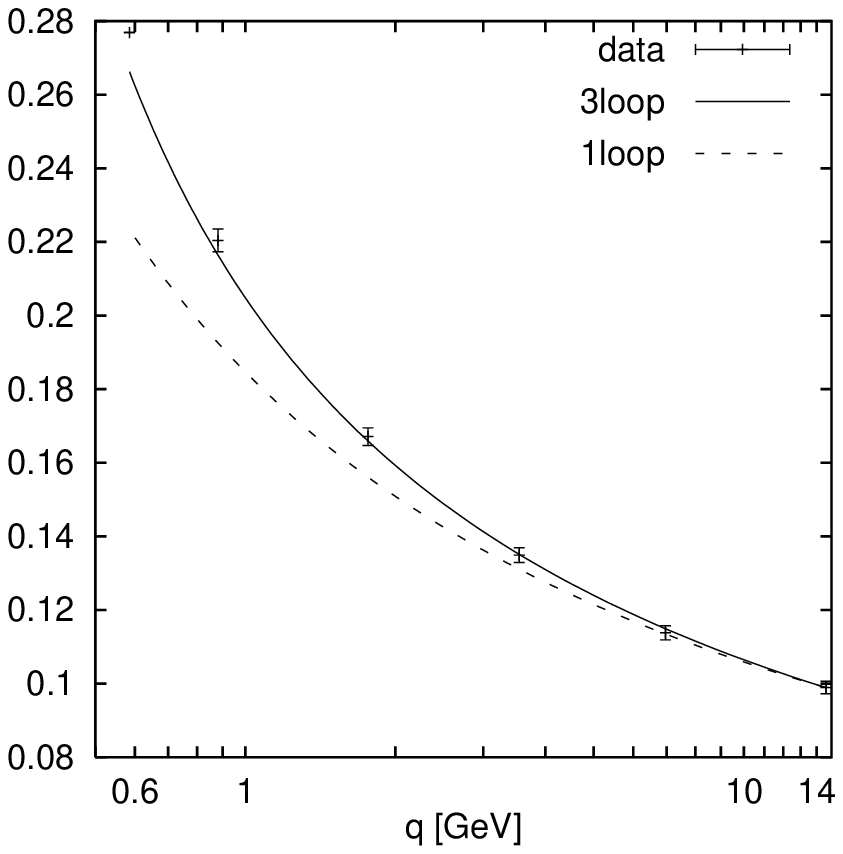}}}
{\it Evolution of the coupling $\aSF$ together with the perturbative
evolution in 1 and 3 loop order.}
\end{minipage} 

\section{Relation to the bare coupling}
The relation of $\aSF$ to the bare coupling $\ao=g_0^2/(4\pi)$ involves
a large scale factor and large coefficients:
\begin{eqnarray}
&& \aSF(  14.06/a)=\ao   +                    4.178(11)\ao^3  \\
&& \aSF(  17.36/a)=\ao \!-\! .36895\ao^2   +  4.042(11)\ao^3 
\end{eqnarray}
Using the tadpole improved coupling $\aot=\ao/P$, we get a better
behaved scale factor, but the coefficients are still relative large.
\begin{eqnarray}
&& \aSF(   1.285/a)   =\aot   +                      1.914(11)\aot^3 \\
&& \aSF(   1.586/a)   =\aot \!-\! .36895\aot^2   +   1.778(11)\aot^3 
\label{aSF_aot_II}
\end{eqnarray}
As an example we can insert $\aot=0.12$ ($\beta=6.5=6/g_0^2
,P=0.6384$) and get in eqn. (\ref{aSF_aot_II}) a 1-loop effect of
4.5\% and a 2-loop effect of 2.6\%. 
This casts doubt on the accuracy of tadpole improvement in this region.
\section{Conclusion}
The present two loop lattice calculation completes the connection
between $\aSF$ and $\aMS$ in the $SU(3)$ gauge theory. It has lead to a
reduction of systematic errors comparable to the case of $SU(2)$.

\section{Acknowledments}
I am grateful to Peter Weisz for providing checks. 
I wish to thank Ulli Wolff for guidance during the
calculation and helpful comments on the manuscript.

\end{document}